\documentclass[aps,prx,twocolumn,longbibliography, footinbib]{revtex4-2}

\usepackage{soul}
\usepackage[dvipsnames]{xcolor}
\usepackage{amsmath}
\usepackage{amsfonts}
\usepackage{graphicx}

\usepackage{amssymb}
\usepackage{amsbsy}
\usepackage{color}

\newcommand{\bra}[1]{\langle{#1}|}
\newcommand{\ket}[1]{|{#1}\rangle}

\newcommand{\beq}{\begin{equation}}
\newcommand{\eeq}{\end{equation}}
\newcommand{\bea}{\begin{eqnarray}}
\newcommand{\eea}{\end{eqnarray}}

\usepackage{hyperref}
\hypersetup{
	colorlinks=true,
	citecolor=blue,
	linkcolor=blue,
	urlcolor=blue}

\begin{document}

\title{Transient and steady-state chaos in dissipative quantum systems}
\author{Debabrata Mondal$^1$, Lea F. Santos$^{2}$, S. Sinha$^1$}
\affiliation{$^1$Indian Institute of Science Education and Research-Kolkata, Mohanpur, Nadia-741246, India\\
$^2$Department of Physics, University of Connecticut, Storrs, Connecticut 06269, USA \\	
}

\begin{abstract}
Dissipative quantum chaos plays a central role in the characterization and control of information scrambling, non-unitary evolution, and thermalization, but it still lacks a precise definition. The Grobe-Haake-Sommers conjecture, which links Ginibre level repulsion to classical chaotic dynamics, was recently shown to fail [Phys. Rev. Lett. {\bf 133}, 240404 (2024)]. We properly restore the quantum-classical correspondence through a dynamical approach based on the von Neumann entropy (VNE) and out-of-time-order correlators (OTOCs), which reveal signatures of chaos beyond spectral statistics. Focusing on the open anisotropic Dicke model, we identify two distinct regimes: transient chaos, marked by rapid early-time growth of VNE and OTOCs followed by low saturation values, and steady-state chaos, characterized by high long-time values. We introduce a random matrix toy model and show that Ginibre spectral statistics signals short-time chaos rather than steady-state chaos. Our results establish VNE dynamics and OTOCs as reliable diagnostics of dissipative quantum chaos across different timescales.
\end{abstract}
\maketitle

Quantum chaos is connected to a wide range of theoretical and experimental phenomena, from thermalization~\cite{Borgonovi2016,Alessio2016} and the failure of many-body
localization~\cite{Suntajs2020, brighi2025}, to the scrambling of quantum
information~\cite{Landsman2019,Xu2024} and the exponential growth of perturbations near black hole horizons~\cite{Maldacena2016JHEP}. While chaos in closed quantum systems is typically characterized by random matrix theory~\cite{Casati1980,Bohigas1984}, including Wigner-Dyson level statistics~\cite{MehtaBook,HaakeBook,WimbergerBook} and eigenstates resembling random vectors~\cite{Berry1977,Borgonovi2016}, defining chaos in open quantum systems remains a challenge \cite{Casati_Dissipative_Chaos_2005, Dahan2022, braun2001}. Dissipation alters dynamical behavior, raising questions about the quantum-classical  correspondence in non-unitary settings.

Efforts to define quantum chaos in open systems have led to the Grobe-Haake-Sommer conjecture~\cite{Grobe1988,Grobe1989,HaakeBook, Akemann2019,Sa2020,Ueda2020,Denisov2019,Sa2020JPhysA,Dukelsky2022, GarciaGarcia2022,Kawabata2023,Prasad2022, wold2025}, which relates classical chaotic attractors with Ginibre level repulsion in the spectrum of the Liouvillian superoperator. However, recent studies have shown that this correspondence can fail~\cite{Lea2024,Minganti2025, DMondal2025, Robnik_2025}. In particular, Ginibre spectral statistics may emerge even when the long-time dynamics is regular~\cite{Lea2024}. This discrepancy suggests that spectral indicators alone may be insufficient to capture the full nature of dissipative quantum chaos and dynamical properties should also be considered~\cite{Minganti2025}. 

%%%%%%%%%%%%%%%%
%%%%%%%%%%%%%%%%%%%%% Schematic %%%%%%%%%%%%%%%%%
\begin{figure}[b]
\includegraphics[width=\columnwidth]{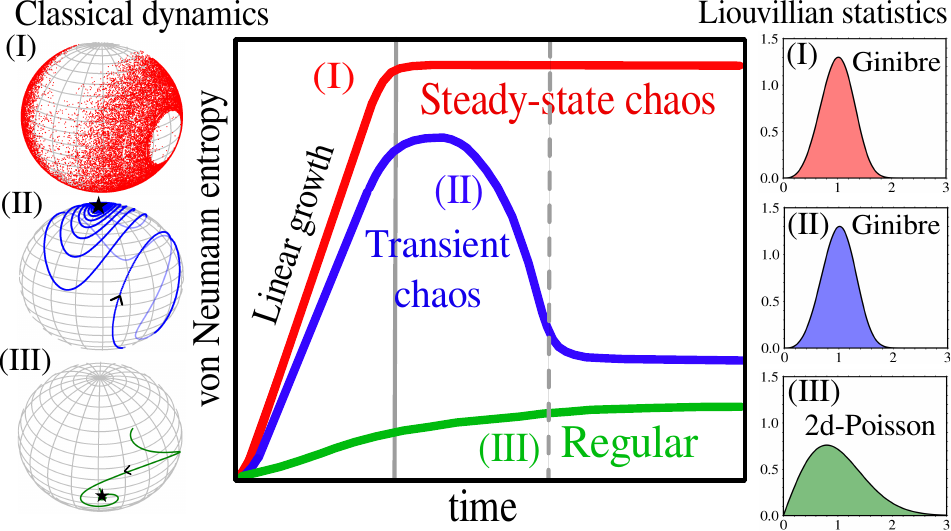}
\caption{Schematic illustration. A dynamical framework restores the quantum-classical correspondence by identifying three regimes: (I) {\em steady-state chaos}, featuring a chaotic attractor in the classical dynamics (left), and linear growth followed by large saturation of the von Neumann entropy (middle panel); (II) {\em transient chaos}, characterized by a regular attractor, and rapid short-time VNE growth followed by decay at long times; and (III) {\em regular dynamics}, showing slow VNE growth and low saturation. Regimes (I) and (II) exhibit Ginibre Liouvillian spectral statistics (right panels), while (III) corresponds to 2d-Poisson statistics.}
\label{fig1}
\end{figure}
%%%%%%%%%%%%%%%%%%%%%%%%%%%%%%%%%%%%%%%%%%%%%%%%%%%%%

In this work, we adopt a dynamical perspective and demonstrate that the quantum-classical correspondence can be restored when chaos is diagnosed through the evolution of the von Neumann entropy (VNE) and out-of-time-order correlators (OTOCs). We generalize fidelity OTOCs (FOTOCs) for non-unitary evolution, making them applicable to open quantum systems. Unlike closed Hamiltonian systems, where rapid early-time mixing often indicates persistent chaotic dynamics, dissipation can suppress long-time chaos, despite signs of chaotic behavior at short times.
As illustrated schematically in Fig.~\ref{fig1}, we identify three regimes with qualitatively distinct behavior: (I) {\em steady-state chaos}, marked by fast initial growth and large long-time saturation values of VNE (middle panel), in agreement with the presence of a chaotic attractor in the classical dynamics (see Bloch sphere on the left side); (II) {\em transient chaos}, which also features rapid early-time scrambling but evolves into a low-entropy steady state, consistent with the presence of a stable attractor; and (III) {\em regular regime}, with slow dynamics and no chaos at any time. 
The FOTOCs exhibit the same qualitative dynamical behavior across the three regimes, showing rapid early-time growth in both cases (I) and (II). Importantly, the right panels of Fig.~\ref{fig1} show that both transient and steady-state chaos exhibit Ginibre level statistics in the Liouvillian spectrum, demonstrating the limitations of spectral features in fully characterizing dissipative quantum chaos.

We explore these ideas using the anisotropic Dicke model (ADM) \cite{Dicke1954,Zhiqiang2017, ScottPerkins2014, Esslinger2021, EmaryBrandes2003, Altland2012, Ciuti2014,Chitra2018, Stitely2020,ScottPerkins2020,Chavez2019,Pilatowsky2020,Pilatowsky2021NatCommun,Villasenor2024ARXIV,SRay2022,DMondal_ADD_2025, tong2025,DasPragna2023a,DasPragna2023b,DasPragna2025,Villasenor2024,pawar2025} in the presence of photon loss. We show that, unlike Liouvillian spectral statistics~\cite{Lea2024}, the dynamics of the VNE of a subsystem of the Dicke model and FOTOCs distinguish transient from steady-state chaos, in agreement with classical diagnostics based on Lyapunov exponents.

To support and generalize our findings, we introduce a random matrix toy model with a tunable Liouvillian. We show that Ginibre spectral statistics emerge whenever the VNE exhibits rapid initial linear growth, regardless of whether it saturates at high values (steady-state chaos) or decays to low values (transient chaos). This analysis reveals that  Liouvillian spectral statistics can reflect early-time chaotic behavior, but does not distinguish different asymptotic regimes.

{\em Model.--} 
The Hamiltonian of the ADM is given by
%\begin{eqnarray}
$$
	\hat{\mathcal{H}} \!= \omega \hat{a}^{\dagger}a \!+\!\omega_0 \hat{S}_{z} \!+\!\frac{\lambda_{-}}{\sqrt{2S}}(\hat{a}\,\hat{S}_{+} \!+\!\hat{a}^{\dagger}\hat{S}_{-}) 
    \!+\!\frac{\lambda_{+}}{\sqrt{2S}}(\hat{a}\hat{S}_{-} \!+\!\hat{a}^{\dagger} \hat{S}_{+}),
    $$
%\end{eqnarray}
%%%%%%%%%%%%%%%%%%%%%%%%
%\begin{eqnarray}
%	\hat{\mathcal{H}} = \omega \hat{a}^{\dagger}a+\omega_0 \hat{S}_{z}+\frac{\lambda_{-}}{\sqrt{\scriptstyle\text{2S}}}(\hat{a}\hat{S}_{+}+\hat{a}^{\dagger}\hat{S}_{-})\nonumber+\frac{\lambda_{+}}{\sqrt{\scriptstyle\text{2S}}}(\hat{a}\hat{S}_{-}+\hat{a}^{\dagger}\hat{S}_{+}),
%\end{eqnarray}
%%%%%%%%%%%%%%%%%%%%%%%%
where $\hbar= 1$, $\hat{a}$ ($\hat{a}^{\dagger}$) annihilates (creates) a cavity photon mode with frequency $\omega$, the collective pseudospin operators $\hat{S}_{z,+,-}$ describe the joint behavior of $N$ two-level atoms with energy splitting $\omega_0$, and $\lambda_{\pm}$ are the atom-photon coupling strengths. In all figures, we set $\omega=1.0$, $\omega_0=1.0$, and $S = N/2=5$ (results for larger $S$ are in the Supplemental Material (SM) \cite{footSM}).

Realistically, photon leakage from the cavity is inevitable,
resulting in the non-unitary time evolution of the density
matrix described by the Lindblad master equation~\cite{Gorini1976, Lindblad1976, Breuer2007},
\begin{eqnarray}
	\frac{d\hat{\rho}}{dt} = \hat{{\cal L}}[\hat{\rho}]= -i[\hat{\mathcal{H}},\hat{\rho}]+\kappa\left[2\hat{a}\hat{\rho }\hat{a}^{\dagger}-\left\{\hat{a}^{\dagger}\hat{a},\hat{\rho}\right\}\right],
	\label{ME}
\end{eqnarray}
where $\kappa$ sets the photon loss rate and $\hat{{\cal L}}$ is the Liouvillian superoperator.

The classical dynamics is derived from the quantum evolution equation, $d\langle \hat{O} \rangle/dt = 
{\rm Tr}(\hat{O} \dot{\hat{\rho}})$, in the limit of a large collective spin, $S\rightarrow \infty$. A phase diagram of the asymptotic classical dynamics of the open ADM was obtained in~\cite{Stitely2020} by analyzing the stability of the fixed points of the equations of motion. The system exhibits a variety of dynamical phases, including normal, superradiant, limit-cycle, and chaotic regimes (see SM~\cite{footSM}).

We focus on the chaotic regime and investigate how classical chaos manifests in the dissipative quantum system.  Our goal is to identify unequivocal quantum signatures of the onset of classical chaos.

%@@@@@@@@@@@@ Quantum dynamics @@@@@@@@@@@@@@@@@@@

{\it Steady-state chaos.--} In the open classical ADM, chaotic dynamics emerges when the regular attractors become unstable. In the quantum domain, the steady state is defined by the zero eigenvalue of the Liouvillian superoperator~\footnote{Despite the presence of a limit cycle in the open classical systems, its quantum analogue can only be probed in the asymptotic limit $S \to \infty$, where some eigenvalues of the Liouvillian become purely imaginary~\cite{Iemini2018,Souza2023,Dutta2025}.}. We show that this long-time quantum state can retain signatures of the underlying classical chaos, a phenomenon we refer to as steady-state chaos.

%%%%%%%%%%%%%%%%%%%%%
We investigate the time evolution of the total VNE defined as 
$\mathcal{S}^{\rm VN} (t)= \mathcal{S}_{\rm spin}^{\rm VN} (t)+\mathcal{S}_{\rm photon}^{\rm VN} (t)$, where $\mathcal{S}_{\rm spin}^{\rm VN}$  ($\mathcal{S}_{\rm photon}^{\rm VN}$) is the VNE of the reduced density matrix of the spin (photon) subsystem obtained by tracing out the photon (spin) degrees of freedom. In the absence of dissipation, these entropies reduce to the entanglement entropy, which builds up dynamically due to the atom–photon interaction and is commonly used to diagnose chaos in isolated systems~\cite{Alessio2016,Borgonovi2016, Piga2019, Wang2004, Fujisaki2003, Ray2016}.
The open system is initially prepared in a product state consisting of a spin coherent state and a photon coherent state \cite{Radcliffe1971}.

Contrary to Hamiltonian systems, where the long-time saturation value of the VNE of subsystems depends on the initial state, dissipation drives the system toward a unique asymptotic behavior. In dissipative dynamics, only the short-time evolution of the VNE depends on the initial state.

%%%%%%%%%%%%%%%%%%%%% Lyapunov VS Sen %%%%%%%%%%
\begin{figure}
   \includegraphics[width=\columnwidth]{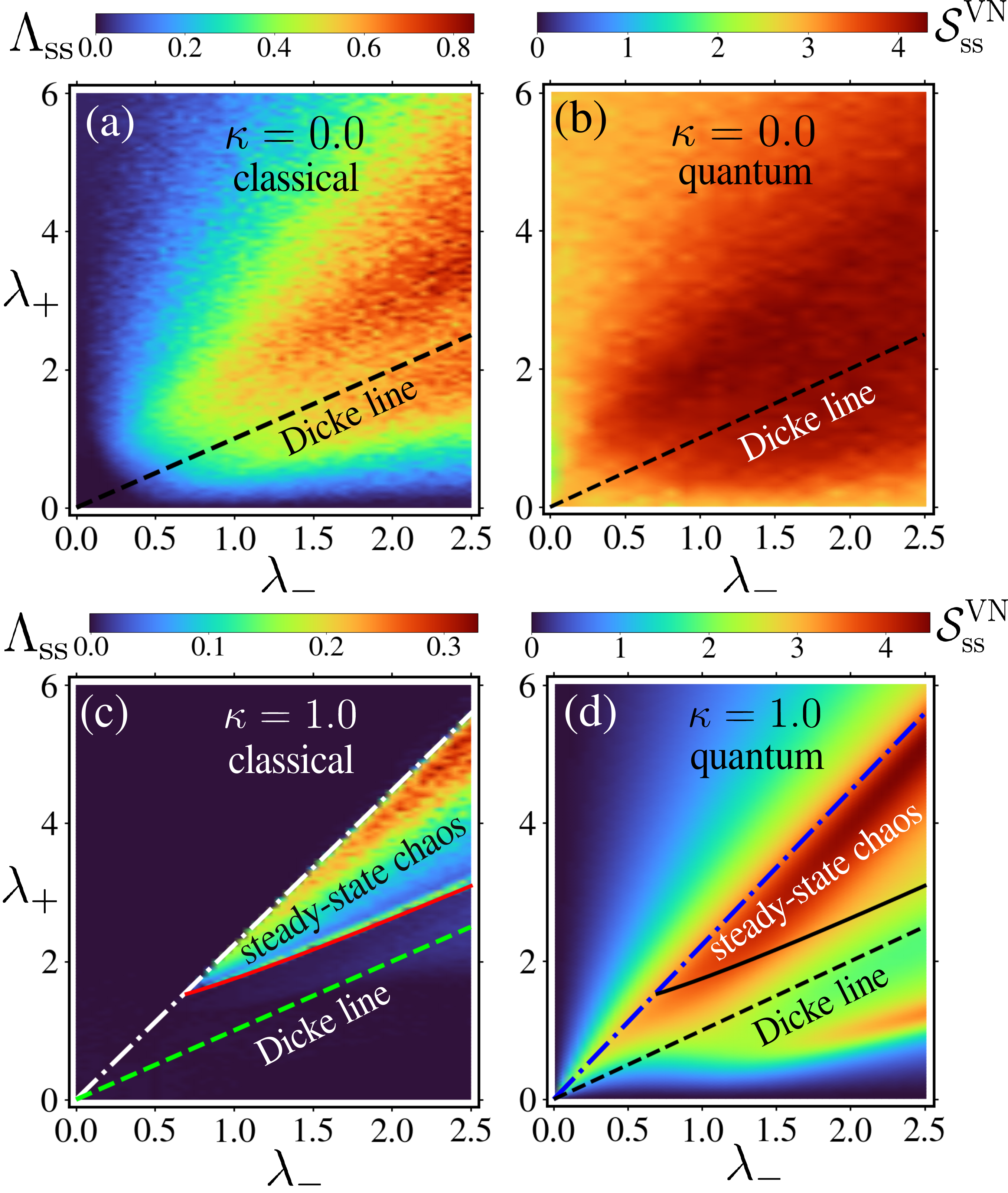}
    \caption{Signature of steady-state chaos on the $\lambda_--\lambda_+$ plane. (a), (c) Averaged long-time Lyapunov exponent $\Lambda_{\rm ss}$ and (b), (d) total von Neumann entropy of the spin and photon subsystems $\mathcal{S}_{\rm ss}^{\rm VN}$ in the (a)-(b) absence of dissipation, $\kappa= 0$, and  (c)-(d) presence of dissipation, $\kappa= 1$.  
    %We neglect the negative values of Lyapunov exponent $\Lambda_{\rm ss}$ and set the minimum value of the color scale to zero.
    }
	\label{fig2}
\end{figure}
%%%%%%%%%%%%%%%%%%%%%%%%%%%%%%%%%%%%%%%%%%%%%%%%%%%%%

The color plots in Fig.~\ref{fig2} compare the average long-time Lyapunov exponent \cite{Strogatz2007,Lichtenberg1992,Hirsch2016}, denoted by $\Lambda_{\rm ss}$ [Figs.~\ref{fig2}(a),\ref{fig2}(c)], with the steady-state total VNE, $\mathcal{S}_{\rm ss}^{\rm VN}$, averaged over an ensemble of initial states [Figs.~\ref{fig2}(b),\ref{fig2}(d)] for the isolated [Figs.~\ref{fig2}(a)-(b)] and dissipative [Figs.~\ref{fig2}(c)-(d)] cases. In the absence of dissipation, the system exhibits large values of $\Lambda_{\rm ss}$ and $\mathcal{S}_{\rm ss}^{\rm VN}$ over a broad region of the coupling parameter space, especially near the Dicke limit, $\lambda_-=\lambda_+=\lambda$, where the classical dynamics is maximally chaotic. When dissipation is introduced, the extent of the chaotic region is significantly reduced. As seen in Figs.~\ref{fig2}(c)-(d), large values of $\mathcal{S}_{\rm ss}^{\rm VN}$ persist only within a narrow triangular region of the parameter space, where chaos is identified classically by positive values of $\Lambda_{\rm ss}$. Outside this region, the classical system becomes regular [see Fig.~\ref{fig2}(c)], including the area around the Dicke limit, and in the quantum case, small but nonzero values of $\mathcal{S}_{\rm ss}^{\rm VN}$ remain visible [Fig.~\ref{fig2}(d)], due to quantum fluctuations at finite spin $S$ and the proximity to a non-equilibrium phase transition. Although the magnitude of $\mathcal{S}_{\rm ss}^{\rm VN}$ does not quantify the absolute degree of chaos, as the Lyapunov exponent does, it efficiently distinguishes between chaotic and regular regimes, with this distinction becoming more pronounced as $S$ increases (see \cite{footSM}). 

The key result in Fig.~\ref{fig2} is the strong quantum-classical correspondence observed not only in the isolated regime but also in the presence of dissipation. The saturation value of the VNE closely follows the classical Lyapunov exponent, as shown by the close agreement between Figs.~\ref{fig2}(c) and \ref{fig2}(d). This demonstrates that, unlike Liouvillian level statistics, which can present Ginibre spectral correlations when the open classical model is regular~\cite{Lea2024}, the steady-state VNE provides a reliable quantum signature of dissipative classical chaos.

%%%%%%%%%%%%%%%%%%% FOTOC %%%%%%%%%%%%%%%%%%%

In isolated systems, the OTOC has also been extensively studied as a quantum diagnostic of chaos \cite{Shenker2015, Shenker2014, Maldacena2016, Swingle2016, Yoshida2016, Yoshii2017, Galitski2017,  SachdevSwingle2017, Fazio2018, 
A_M_Rey2018,Ray2018,GarciaMata2018, Lewis-Swan2019, Chavez2019, Pilatowsky2020, Chavez2023, Lakshminarayan2020,Tian2022}. It is defined as $\mathcal{F}(t) = \langle  \hat{W}^{\dagger}(t)\hat{V}^{\dagger}(0) \hat{W}(t)\hat{V}(0)\rangle$, where the operator $\hat{W}$ evolves in time and the operator $\hat{V}$ is fixed at $t=0$. The OTOC quantifies the spread of quantum information and sensitivity to perturbations. A variant of this correlator is the FOTOC~\cite{Lewis-Swan2019} obtained by choosing $\hat{V}=\hat{\rho}(0)$ and $\hat{W}=e^{i\delta\phi\hat{G}}$, where $\delta \phi$ is a small perturbation and $\hat{G}$ is a Hermitian operator. In the perturbative limit $\delta \phi \ll 1$, the FOTOC can be written in terms of the variance  $(\Delta{G}(t))^2 = \langle \hat{G}^2 (t)\rangle - \langle \hat{G} (t)\rangle^2$ as  $\mathcal{F}_{G} (t) \approx 1-\delta\phi^2 (\Delta{G} (t))^2$, and its early-time exponential growth indicates instability and chaos.

We generalize the FOTOC to open quantum systems by evolving the operator $\hat{W}$ under non-unitary dynamics according to $d\hat{W}/dt = \mathcal{L}^{\dagger}[\hat{W}]$. 
Similar to the VNE, the steady-state value of the standard deviation $(\Delta G)_{\rm ss}$ obtained from the FOTOC also serves as an indicator of steady-state chaos (see SM~\cite{footSM}).

%%%%%%%%%%%%%%%%%%%%% Transient chaos %%%%%%%%%%%%%%

{\it Transient chaos.--} Complementing the long-time analysis, we examine the short-time dynamics of the VNE and $\Delta G(t)$, both of which capture signatures of transient chaos. In isolated classical systems, chaos arises from rapid mixing and sensitivity to initial conditions that emerge early in the evolution. When dissipation is introduced, the system may still display chaotic behavior at short times, even if the long-time dynamics become regular. This transient chaos reflects a competition between the early-time chaotic dynamics inherited from the isolated system and the eventual convergence to a regular attractor of the dissipative system. In the quantum regime, a similar phenomenon occurs: the dynamics can ultimately settle into a non-chaotic steady state, even if remnants of chaos, such as rapid entropy growth, are still visible at early times. In our system, the Dicke limit offers a natural setting to explore this behavior, as it exhibits maximal chaos in the absence of dissipation, while increasing the dissipation rate $\kappa$ progressively suppresses chaos at long times, as the superradiant fixed point becomes a stable attractor.

\begin{figure}
	\includegraphics[width=\columnwidth]{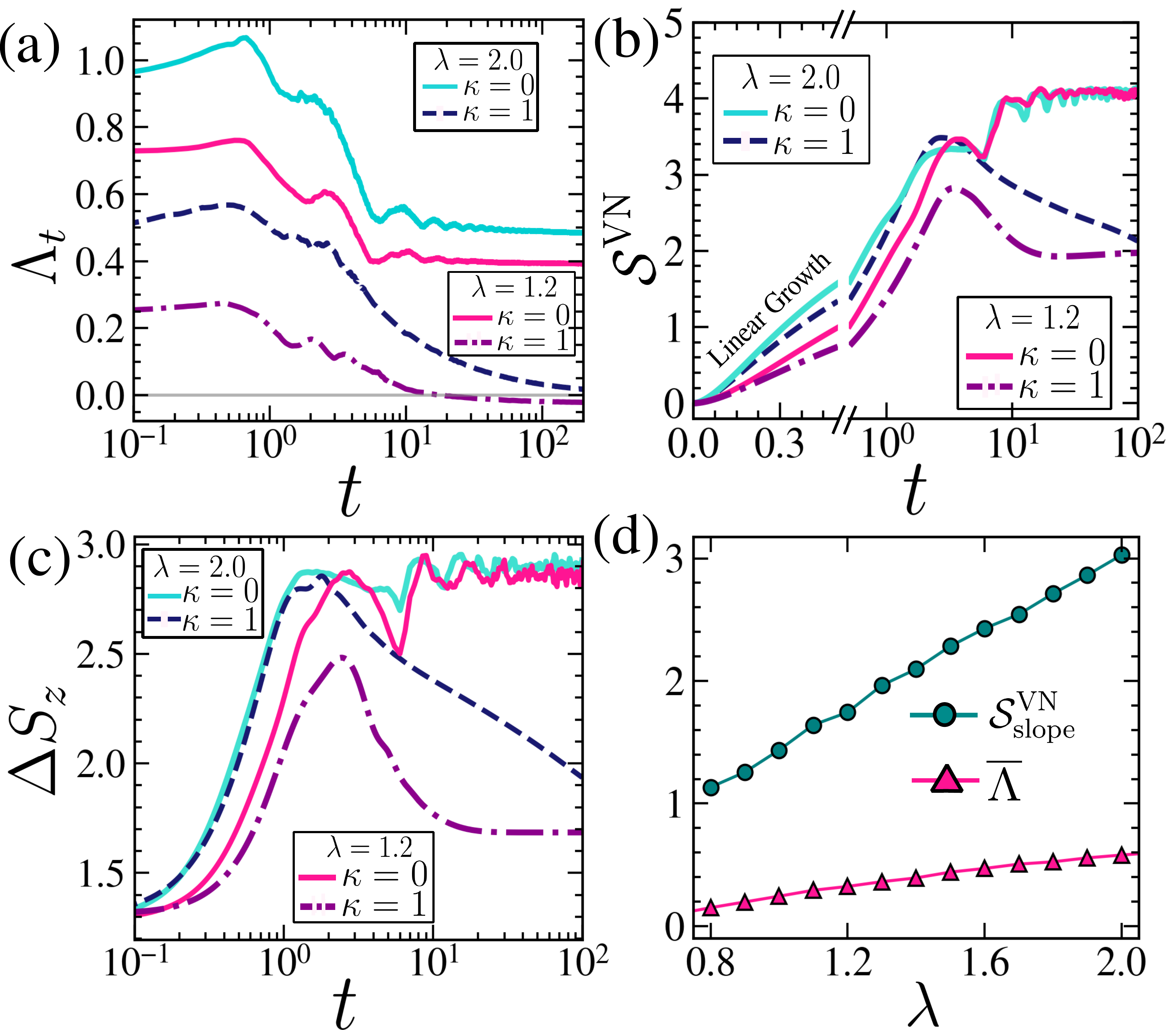}
	\caption{Transient chaos vs steady-state chaos in the Dicke limit $\lambda_-=\lambda_+=\lambda$. (a) Finite-time ensemble averaged Lyapunov exponent $\Lambda_t$, (b) total VNE $\mathcal{S}^{\rm VN}(t)$, and (c) standard deviation $\Delta S_z(t)\approx \sqrt{1-\mathcal{F}_z(t)}/\delta\phi$ obtained from FOTOC $\mathcal{F}_z(t)$ of $z$ component of spin for $\delta\phi \ll 1/S$,  for atom-photon coupling strengths $\lambda=1.2, 2.0$ and  $\kappa = 0, 1$. In (b), the time axis is linear for $t\le 0.5$ and logarithmic for $t>0.5$.  (d) The linear growth rate of VNE, $\mathcal{S}^{\rm VN}_{\rm slope}$, and the time-averaged Lyapunov exponent, $\overline{\Lambda}$, within a time interval $t\in[0, 0.5]$ as a function of the coupling strength $\lambda$ for $\kappa = 1$. 
    %The inset of (d) shows the initial linear growth of $\mathcal{S}^{\rm en}(t)$ for $\lambda=1.5$ and $\kappa = 1$. 
    }
	\label{fig3}
\end{figure}
%%%%%%%%%%%%%%%%%%%%%%%%%%%%%%%%%%%%%%%%%%%%%%%%%%%%%

We demonstrate this behavior in Fig.~\ref{fig3}(a), where we consider the Dicke limit and compare the finite-time Lyapunov exponent, $\Lambda_t$, averaged over initial phase-space points for the isolated and dissipative systems.  In both cases, $\Lambda_t$ is initially nonzero, indicating chaotic dynamics at short times. However, in the presence of dissipation, $\Lambda_t$ gradually decays to zero at long times, reflecting the growing influence of the stable attractor and the suppression of chaos. This decay confirms the existence of transient chaos in the dissipative regime.

To uncover quantum signatures of transient chaos in the Dicke limit, we examine the time evolution of the von Neumann entropy $\mathcal{S}^{\rm VN}(t)$ [Fig.~\ref{fig3}(b)] and the standard deviation of the $z$ component of the spin operator, $\Delta S_z(t)$ [Fig.~\ref{fig3}(c)], both averaged over an ensemble of initial states. In both isolated and dissipative settings, these observables exhibit rapid growth at short times, consistent with short-time chaotic dynamics. However, at long times, similar to the behavior of $\Lambda_t$, the asymptotic values of $\mathcal{S}^{\rm VN}(t)$ and $\Delta S_z(t)$ are significantly reduced for the dissipative model, indicating a transition to regular dynamics. This contrast between rapid short-time growth and long-time suppression reflects the presence of transient chaos also for the quantum dissipative system. Similar behavior is also observed for the standard deviation of the photon number.

Further insight into the nature of transient chaos is provided by its dependence on the atom-photon coupling $\lambda$. As seen in Figs.~\ref{fig3}(b)-(c), the short-time growth rates of $\mathcal{S}^{\rm VN}(t)$ and $\Delta S_z(t)$ increase with the atom-photon coupling $\lambda$, in agreement with the larger values of the short-time Lyapunov exponent observed in Fig.~\ref{fig3}(a). This dependence on $\lambda$ is quantified in Fig.~\ref{fig3}(d), where we show that both the growth rate of VNE, $\mathcal{S}^{\rm VN}_{\rm slope}$, and the time-averaged Lyapunov exponent computed over small time windows $t\in [0,0.5]$ scale nearly linearly with $\lambda$. 
%%%%%%%%%%%%%%%%%%
Moreover, Fig.~\ref{fig3}(b) shows that in the Dicke limit, the VNE grows linearly at early times, supporting its role as a sensitive probe of transient chaos.

{\it Random matrix toy model.--} To further elucidate the distinction between transient and steady-state chaos in open quantum systems, and to investigate the connection with the Liouvillian spectral statistics, we introduce a random matrix toy model governed by a tunable Liouvillian superoperator. 

The total Hamiltonian is given by $\hat{\mathcal{H}} = \hat{\mathcal{H}}_{\rm TD} + \frac{\mu}{\sqrt{N}} \hat{\mathcal{H}}_{\rm I}$, where $\hat{\mathcal{H}}_{\rm TD}$ is a tridiagonal $N\times N$ random matrix and $\hat{\mathcal{H}}_{\rm I}$ is  a perturbation whose structure we vary to explore different dynamical regimes. The parameter $\mu \in [0,1]$ controls the relative strength of the perturbation. Dissipation is incorporated through a random jump operator $\hat{L}$, which has nonzero real random entries only on the first subdiagonal, $\hat{L}_{m+1,m}\neq 0$. The resulting Liouvillian superoperator takes the standard Lindblad form,
%%%%%%%%%%%%%%%%%%%%%%%%%%%%%%%%%%
\begin{align}
	\hat{\mathcal{L}} = - & \,i  \left[\hat{\mathcal{H}} \otimes  \mathbb{I} - \mathbb{I} \otimes \hat{\mathcal{H}}^*\right] \nonumber \\
	+ \,& \gamma\Big( 2\hat{L} \otimes  \hat{L}^* 
	- \hat{L}^\dagger \hat{L} \otimes \mathbb{I}
	-  \mathbb{I} \otimes  \hat{L}^{\mathrm{T}} \hat{L}^* \Big),
	\label{Liouvillian}
\end{align}
where $\gamma$ is the dissipation strength and ``T" represents the transposition of a matrix.

We first choose $\hat{\mathcal{H}}_{\rm I}=\hat{\mathcal{H}}_{\rm GOE}$, that is, a full random matrix drawn from the Gaussian orthogonal ensemble. In the unitary case ($\gamma=0$), the spectral statistics of $\hat{\mathcal{H}}$ interpolates from Poisson ($\mu=0$) to Wigner-Dyson ($\mu=1$) level spacing distribution. When dissipation is included ($\gamma=1$), the Liouvillian spectrum transitions from the two-dimensional (2d) Poisson ($\mu=0$) to the Ginibre ($\mu=1$) statistics of non-Hermitian random matrices~\cite{HaakeBook,Ueda2020,Denisov2019, Sa2020}, as shown in the SM~\cite{footSM}.

To probe the resulting dynamics of the open system ($\gamma=1$), we initialize it in a pure product state, $\hat{\rho}(0)$, built with the eigenstates of $\hat{\mathcal{H}}_{\rm TD}$, and evolve it under the Liouvillian in Eq.~\eqref{Liouvillian}. We assume that the dimension $N$ of the total Hamiltonian is constructed as the tensor product of two abstract subsystems, each of dimension $M$, so $N=M^2$.

\begin{figure}[h]
	\includegraphics[width=\columnwidth]{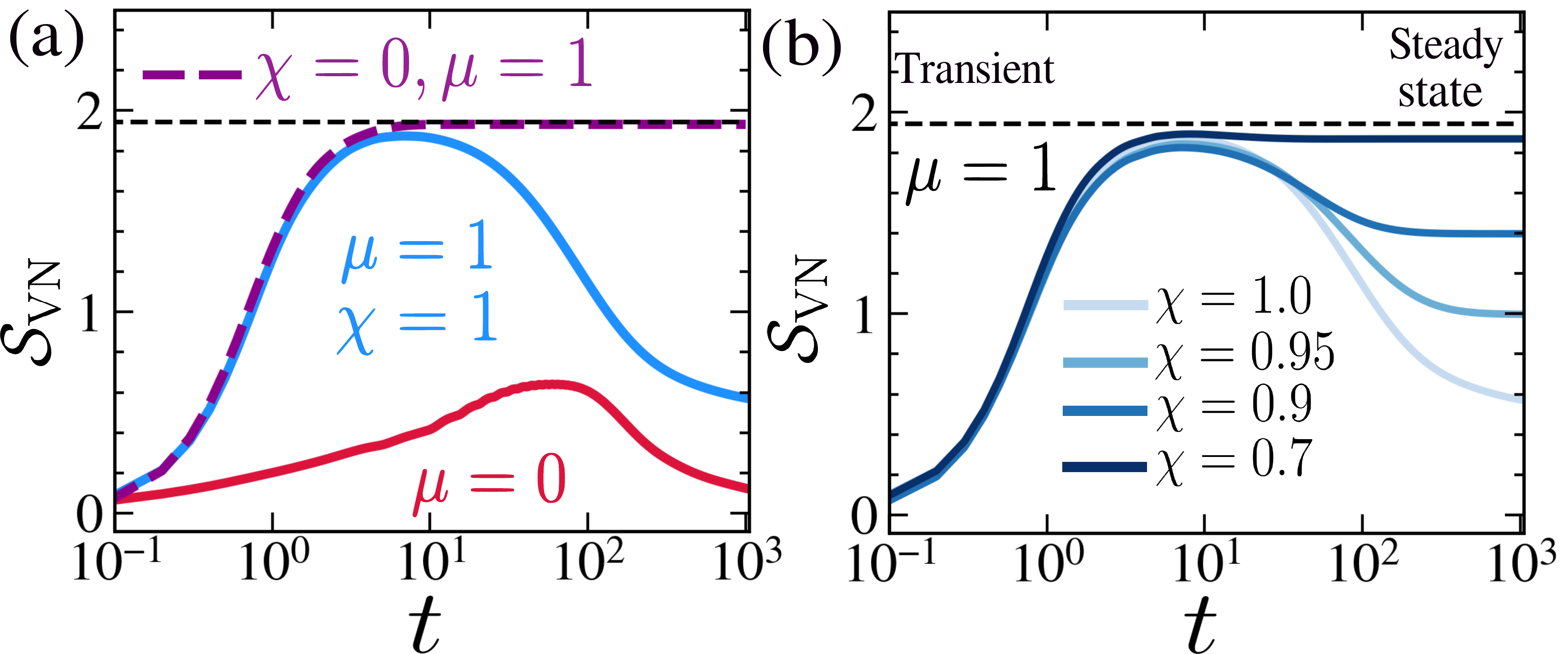}
	\caption{Transient and steady-state chaos for the random matrix toy model. (a) Time evolution of the von Neumann entropy $\mathcal{S}_{\rm VN}$, averaged over an ensemble of random matrices, for the Liouvillian in Eq.~(\ref{Liouvillian}) with $\mu = 0$ (solid red line), resulting in 2d Poisson level statistics, with $\mu = 1$, $\chi=0$ (dashed violet line), resulting in Ginibre spectral statistics, and the projected case with $\mu = 1$, $\chi=1$ (solid blue line), also resulting in Ginibre spectral statistics.  (b) Evolution of $\mathcal{S}_{\rm VN}$ for different values of the deformation parameter $\chi$ and fixed $\mu = 1$. The horizontal dashed line in (a)-(b) indicates the maximal entropy of the subsystem, $\mathcal{S}^{\rm max}_{\rm VN} = \frac{1}{2}\ln(N)$.  We use $N=49$ and $\gamma = 1$. The random numbers are drawn from a Gaussian distribution with zero mean and unit variance.
}
	\label{fig4}
\end{figure}

In the Ginibre regime ($\mu=1$), the dashed line in Fig.~\ref{fig4}(a) shows that the VNE of the reduced density matrix of a subsystem $\mathcal{S}_{\rm VN}$, averaged over an ensemble of random matrices, grows rapidly and saturates near the maximum value $\mathcal{S}^{\rm max}_{\rm VN} = \frac{1}{2}\ln(N)$, indicating both short-time and steady-state chaos.  In contrast, for the 2d Poisson statistics ($\mu=0$), the entropy grows slowly and saturates at a much lower value, consistent with the absence of chaos. The steady-state density matrix in this regime retains a significant degree of purity and can be expressed as $\hat{\rho}_{\rm R} = \sum_i \eta_i \ket{u_i}\bra{u_i}$, where only a few eigenvalues $\eta_i$ are significantly large.

To decouple transient chaos from steady-state behavior, we consider a second choice of $\hat{\mathcal{H}}_{\rm I}$ by deforming the chaotic Hamiltonian via $\hat{\mathcal{H}}_{\rm I} = \hat{\mathcal{P}}(\chi) \hat{\mathcal{H}}_{\rm GOE} \hat{\mathcal{P}}(\chi)$, where 
	$\hat{\mathcal{P}}(\chi) = \mathbb{I}-\chi(\ket{u_1}\bra{u_1}+\ket{u_2}\bra{u_2})$
is a projection operator that protects the dominant eigenstates $\ket{u_1}$ and $\ket{u_2}$ of the regular steady-state density matrix $\hat{\rho}_{\rm R}$ from mixing with the random part. The parameter $\chi \in [0,1]$ controls the strength of this suppression and allows for retention of partial purity in the steady state.

%%%%%%%%%%%%%%%%%%%%% Toy model %%%%%%%%%%%%%%%%%%%%
%%%%%%%%%%%%%%%%%%%%%%%%%%%%%%%%%%%%%%%%%%%%%%%%%%%%%

This deformation preserves the Ginibre spectral statistics at $\mu=1$ for all values of $\chi$ (see SM~\cite{footSM}), and it also leads to the rapid initial growth of the VNE, as shown by the blue solid line in Fig.~\ref{fig4}(a). However, the long-time entropy saturates well below the maximum value $\mathcal{S}^{\rm max}_{\rm VN}$, indicating  absence of chaotic behavior in the steady state. This behavior is consistent with the scenario of transient chaos.

Figure~\ref{fig4}(b) further demonstrates that while the short-time growth of the entropy is largely unaffected by the projector, the long-time saturation value decreases monotonically with increasing $\chi$. This behavior closely mirrors that of the open ADM, where dissipation suppresses long-time chaos but preserves transient chaos. These results show that Ginibre spectral statistics is directly associated with transient chaos, but does not necessarily reflect steady-state chaos. This resolves the perceived breakdown of the quantum-classical correspondence reported in~\cite{Lea2024}.

{\it Conclusions.--} This work demonstrated that chaos in open quantum systems manifests differently across timescales and cannot be fully characterized by spectral statistics alone. Using the open anisotropic Dicke model, we showed that the von Neumann entropy of  subsystems of the model and OTOCs reliably distinguish between transient and steady-state chaos, restoring the quantum-classical correspondence in the presence of dissipation at both short and long times. In contrast, Ginibre spectral statistics of the Liouvillian spectrum are connected to early-time chaotic behavior and therefore cannot differentiate between the two asymptotic regimes, as supported by our random matrix toy model. %\textcolor{blue}{The present work emphasizes that the Liouvillian spectral statistics can only reliably capture the chaos at short time, however, analyzing the properties of the steady state is more important to understand the long time behavior.}

{\it Acknowledgments.--} L.~F.~S. thanks David Villase\~nor and Pablo Barberis-Blostein for useful discussions and support from the National Science Foundation Center for Quantum Dynamics on Modular Quantum Devices (CQD-MQD) under Award Number 2124511.

%%%%%%%%%%%%%%
%REFERENCES
%%%%%%%%%%%%%%

\bibliography{bibliography.bib}

%@*@*@*@*@*@*@*@*@*@*@*@*@*@*@*@*@*@*@*@*@*@*@*@*@*@
%  				SUPPLEMENTARY MATERIAL
%@*@*@*@*@*@*@*@*@*@*@*@*@*@*@*@*@*@*@*@*@*@*@*@*@*@
%\pagebreak

\onecolumngrid

\vspace*{0.4cm}

\begin{center}
	{\large \bf Supplemental Material: 
		\\Transient and steady-state chaos in dissipative quantum systems}\\
	\vspace{0.6cm}
	Debabrata Mondal$^1$, Lea F. Santos$^2$, and S. Sinha$^1$\\
	$^1${\it Indian Institute of Science Education and Research-Kolkata,Mohanpur, Nadia-741246, India}
	
	$^2${\it Department of Physics, University of Connecticut, Storrs, Connecticut 06269, USA}
\end{center}

%\twocolumngrid

\renewcommand{\theequation}{S\arabic{equation}}
\renewcommand{\thefigure}{S\arabic{figure}}

This supplemental material provides additional figures and analyses that support the discussions in the main text. In Sec.~\ref{Sec:Classical_dynamics}, we briefly examine the classical dynamics of the open anisotropic Dicke model (ADM), focusing on the phase diagram of non-equilibrium phases and the emergence of dissipative chaos.  In Sec.~\ref{Sec:Steady_state_chaos}, steady-state chaos is analyzed with the saturation value of the von Neumann entropy (VNE) and standard deviations of the $z$ component of the spin operator and the photon number. Analysis of the dependence on the spin size is also provided.
Finally, Sec.~\ref{Sec:Random_matrix_toy_model} presents results for the spectral statistics of the Liouvillian associated with the random matrix toy model.

\section{Classical dynamics in the open anisotropic Dicke model}
\label{Sec:Classical_dynamics}

%%%%%%%%%%%%%%%%%%%%% Phase diagram %%%%%%%%%%%%%%%%%
\begin{figure}[t]
	\includegraphics[width=0.6\columnwidth]{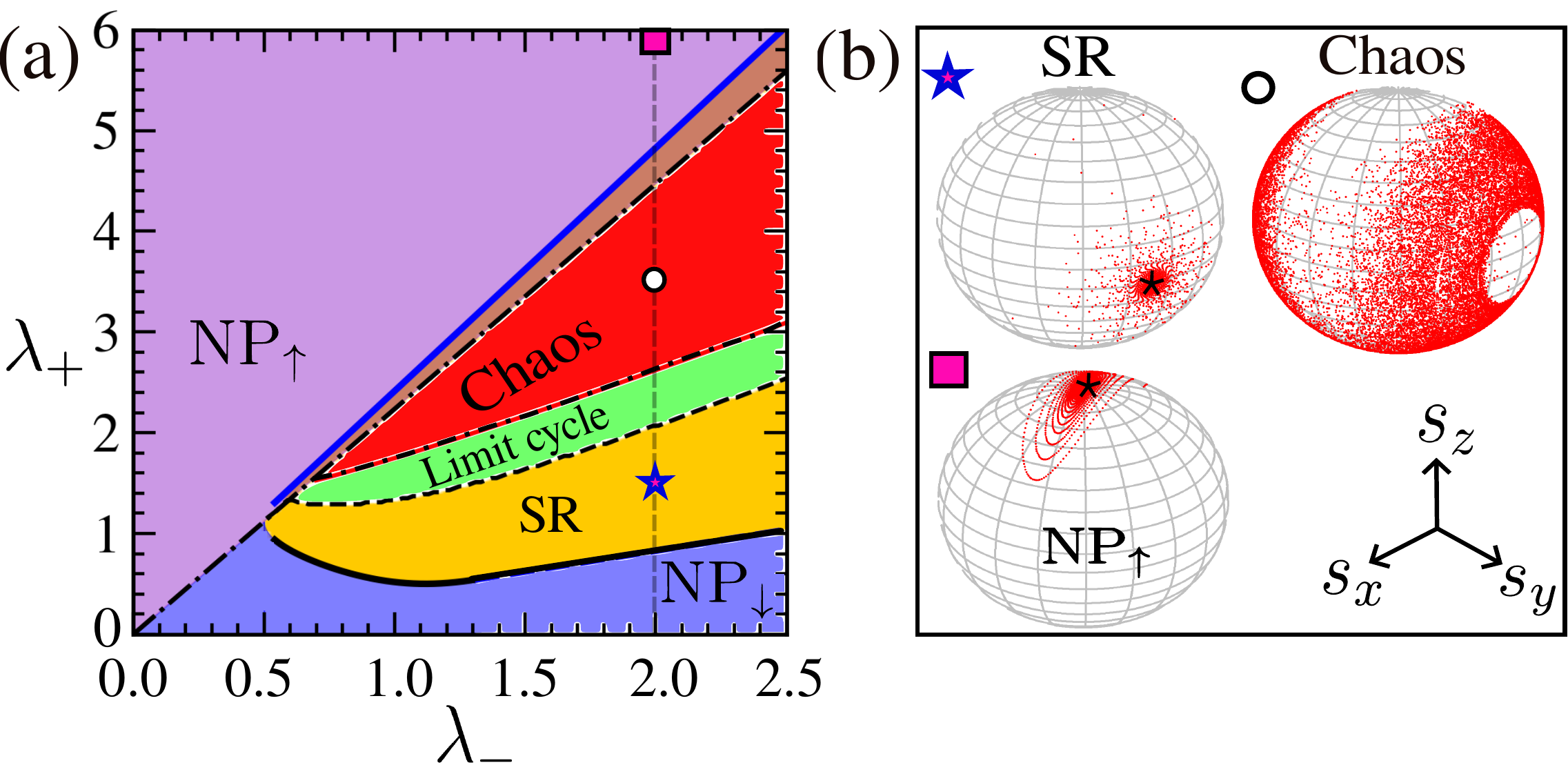}
	\caption{ The classical dynamics of the open anisotropic Dicke model. (a) The classical phase-diagram on the $\lambda_--\lambda_+$ plane. (b) The spin dynamics over the Bloch sphere at different values of $\lambda_{+}$ for $\lambda_{-}=2.0$. All energies (time) are measured by $\omega (1/\omega)$. We set $\omega=1.0,\omega_0=1.0$ and dissipation strength $\kappa = 1$. }
	\label{Fig_Sup1}
\end{figure}
%%%%%%%%%%%%%%%%%%%%%%%%%%%%%%%%%%%%%%%%%%%%%%%%%%%%%

In this section, we discuss the classical dynamics and the various non-equilibrium phases that arise in the ADM in the presence of photon loss.
The non-unitary evolution of the system's density matrix $\hat{\rho}$ is governed by the Lindblad master equation, as given in Eq.~(1) of the main text. The time evolution of the expectation value of any operator $\hat{\mathcal{O}}$ is obtained using the relation 
$$\frac{d\langle \hat{\mathcal{O}} \rangle }{ dt} = 
{\rm Tr}(\hat{\mathcal{O}} \dot{\hat{\rho}}).$$
%%%%%%%%%%%%%%%%%
In the limit $S\rightarrow \infty$, the scaled operators $\hat{a}/\sqrt{S} = (\hat{x} + i \hat{p})/\sqrt{2}$ and $\hat{\vec{s}} = \hat{\vec{S}}/S$ behave classically, since they satisfy the commutation relations $[\hat{x},\hat{p}]= i/S$ and $[\hat{s}_{a}, \hat{s}_{b }] = i \epsilon_{abc} \hat{s}_{c }/S$, where $1/S$ plays the role of a reduced Planck constant.
%%%%%%%%%%%%%%%%%%
The classical equations of motion for the scaled observables can be derived from the master equation and are given by
\begin{subequations}
	\begin{eqnarray}
		\dot{\alpha} &=& -(\kappa+i\omega) \alpha-i(\lambda_-s_-+\lambda_+s_+), \\
		\dot{s}_+ &=& i\omega_0 s_+-i s_{z}(\lambda_-\alpha^*+\lambda_+\alpha),\\
		\dot{s}_{z} &=& -\frac{i}{2}[ \lambda_-(\alpha s_+-\alpha^*s_-)+\lambda_+(\alpha^* s_+-\alpha s_-)],
	\end{eqnarray}
	\label{EOM}
\end{subequations}
where $\alpha = (x+i p)/\sqrt{2}=\sqrt{n}\exp(i\psi)$ represents scaled classical photon field with number $n=|\alpha|^2$ and phase $\psi$. The scaled spin vector can be written as $\vec{s} = (\sin\theta\cos\phi,\sin\theta\sin\phi,\cos\theta)$, where $s_z=\cos\theta$ and $\phi$ are conjugate variables.

To understand the overall dynamics and  describe the different non-equilibrium phases, we analyze the fixed points and attractors of the equations of motion in Eqs.~\eqref{EOM}. The various phases of the above model for the coupling parameters  $\lambda_-$ and $\lambda_+$ and a fixed dissipation strength $\kappa$ are summarized in the phase diagram, in Fig.~\ref{Fig_Sup1}(a). 
%%%%%%%%%%%%%%%%%%%%%%%%

For the ADM, there are two types of normal phases, NP$_{\downarrow}$ and NP$_{\uparrow}$, characterized by vanishing photon number $n^*=0$ and spin polarization $s_{z}^*=-1$ and $s_{z}^*=+1$, respectively. Fixing $\lambda_-=2$ and following the vertical line in  Fig.~\ref{Fig_Sup1}(a) by increasing  $\lambda_+$, the normal phase NP$_{\downarrow}$ undergoes a transition to a superradiant phase (SR) with non-zero photon number $n^*\ne 0$ and spin polarization $|s_z^*|< 1$.
Then the SR phase becomes unstable at a critical coupling and undergoes a Hopf bifurcation, giving rise to a limit cycle (LC). Further increasing $\lambda_+$ leads to the instability of the limit cycle and emergence of chaotic motion, which can be quantified with the Lyapunov exponent \cite{Strogatz2007,Lichtenberg1992,Hirsch2016}. There is a narrow region above the chaotic domain in the phase-diagram, where the SR phase regains its stability and coexists with the stable normal phase NP$_{\uparrow}$.
As the interaction gets even stronger,  the SR phase disappears entirely, and only the normal phase NP$_{\uparrow}$ remains as stable steady state. The detailed description of these non-equilibrium phases and the dissipative dynamics of the open classical ADM is presented in Ref.~\cite{Stitely2020}. The spin dynamics across distinct dynamical regimes for different values of $\lambda_+$  with fixed $\lambda_-$ are illustrated on the Bloch sphere in Fig.~\ref{Fig_Sup1}(b).

\section{Steady-state chaos in the open anisotropic Dicke model}
\label{Sec:Steady_state_chaos}

\begin{figure}[h]
	\includegraphics[width=\columnwidth]{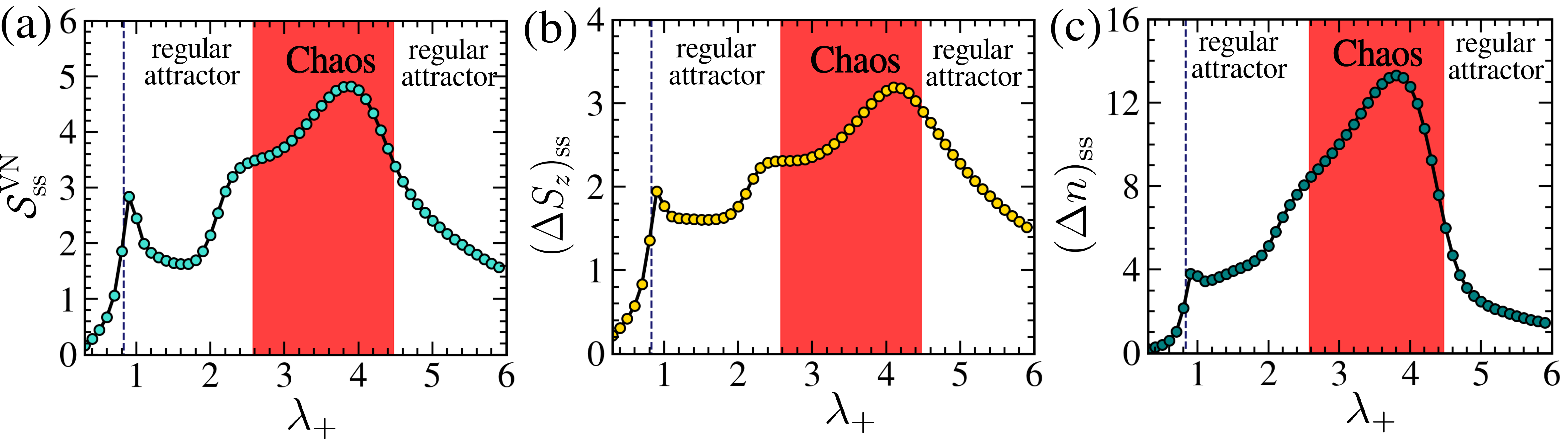}
	\caption{ Signature of steady-state chaos based on the saturation value of the (a) total von Neumann entropy $\mathcal{S}_{\rm ss}^{\rm VN}$,  (b) fluctuation of the $z$ component of the spin operator $(\Delta S_z)_{\rm ss}$, and (c) fluctuation of the photon number  $(\Delta n)_{\rm ss}$ as a function of $\lambda_{+}$ for $\lambda_{-}$ = 2. We consider the parameters:  $\omega = \omega_0 = 1, \kappa = 1, S = 5$.}
	\label{Fig_Sup2}
\end{figure}

As shown in Fig.~\ref{Fig_Sup1}(a), the open ADM exhibits chaos within a narrow triangular region of the phase-diagram, where no stable attractor persists. The presence of chaos can be quantified using the saturation value of the classical Lyapunov exponent $\Lambda_{\rm ss}$, as shown in Fig.~2(c) of the main text. Quantum mechanically, such steady-state chaos can be captured by the saturation value of the total VNE, $\mathcal{S}_{\rm ss}^{\rm VN}$, and the standard deviations of physical quantities, such as the $z$ component of the spin operator, $(\Delta S_z)_{\rm ss}$, and the photon number, $(\Delta n)_{\rm ss}$,  extracted from the corresponding FOTOCs. For a fixed coupling strength $\lambda_{-}=2$, we examine in Fig.~\ref{Fig_Sup2} how the values of these three quantities change as $\lambda_{+}$ increases, corresponding to the vertical line in Fig.~\ref{Fig_Sup1}(a). We see that all three quantities reach their maximum values in the chaotic regime.

%%%%%%%%%%%%%%%%%%%%% Steady-state chaos %%%%%%%%%%%%%
%%%%%%%%%%%%%%%%%%%%%%%%%%%%%%%%%%%%%%%%%%%%%%%%%%%%%

\subsection{Dependence on the spin size}

\begin{figure}[b]
	\includegraphics[width=0.45\columnwidth]{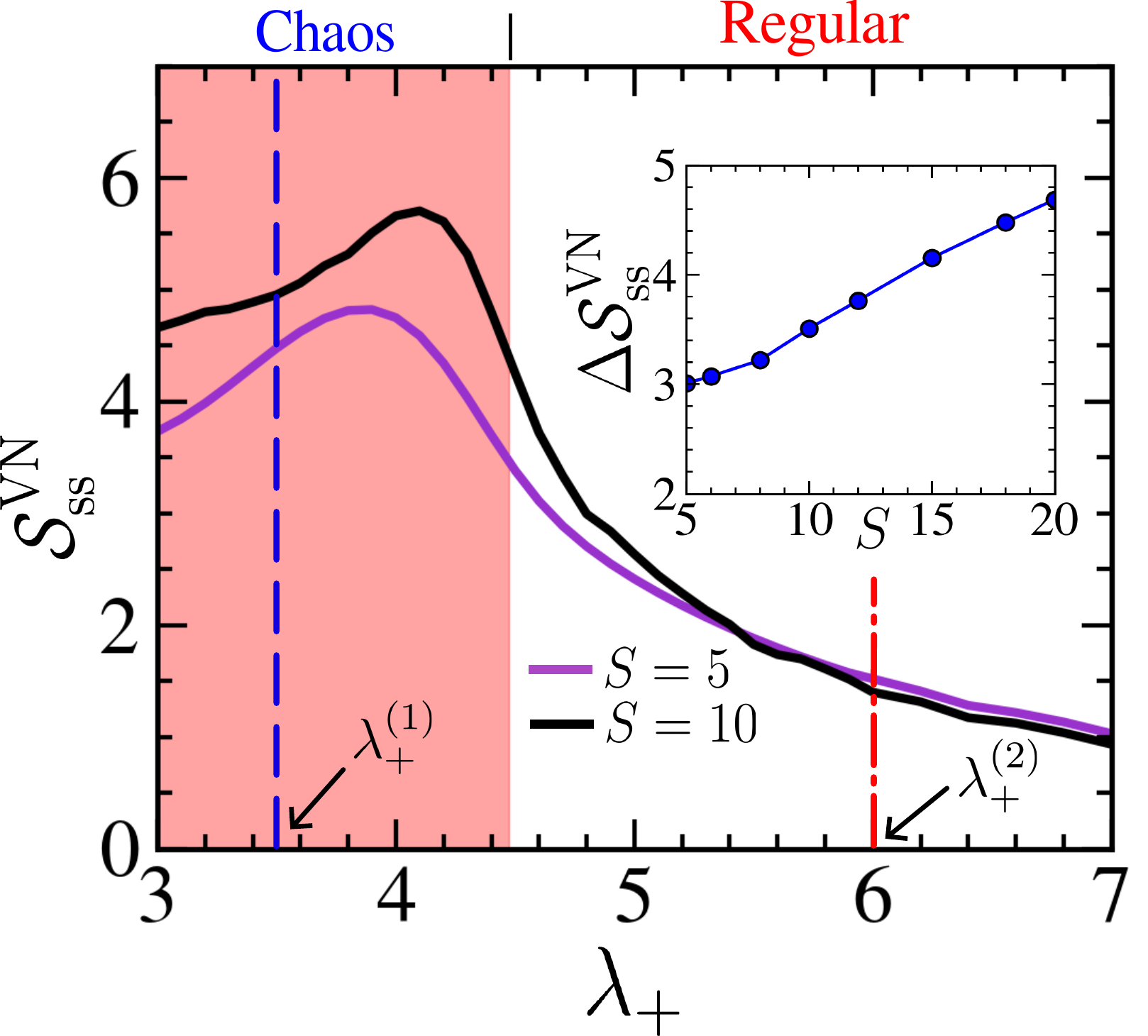}
	\caption{Crossover from the chaotic to the regular regime. The main panel shows how the steady state value of total VNE, $\mathcal{S}^{\rm VN}_{\rm ss}$, changes with the coupling $\lambda_+$ for spin magnitudes $S = 5$ and $10$. The red shaded area indicates the chaotic regime. The inset shows the difference $\Delta\mathcal{S}^{\rm VN}_{\rm ss}$ in the values of the total VNE at the couplings $\lambda_{+}^{(1)}$ (chaotic region) and $\lambda_{+}^{(2)}$ (regular region) as a function of spin magnitude $S$. Chosen parameters: $\omega = \omega_0 = 1, \lambda_-=2, \kappa = 1$.}
	\label{Fig_Sup3}
\end{figure}

As discussed in the main text, steady-state chaos in the open quantum ADM is effectively captured by the relative enhancement of the subsystem von Neumann entropy, $\mathcal{S}^{\rm VN}_{\rm ss}$, as illustrated in the color-scale phase diagram of Fig.2~(d) of the main text. Here, we focus on the suppression of chaos associated with the emergence of the stable normal phase NP$_{\uparrow}$ as the coupling strength $\lambda_+$ increases. Figure~\ref{Fig_Sup3} shows that $\mathcal{S}^{\rm VN}_{\rm ss}$ decreases gradually with increasing $\lambda_+$, signaling a crossover from the chaotic to the regular regime. The relative difference in VNE between these regimes grows with the spin magnitude $S$, yielding a sharper quantum distinction between chaotic and regular dynamics.

To quantify this contrast, we define  
$$\Delta\mathcal{S}^{\rm VN}_{\rm ss} = \mathcal{S}^{\rm VN}_{\rm ss} (\lambda_{+}^{(1)})-\mathcal{S}^{\rm VN}_{\rm ss} (\lambda_{+}^{(2)}),$$ 
where the coupling strengths $\lambda_{+}^{(1)}$ and $\lambda_{+}^{(2)}$ lie in the chaotic and regular regions, respectively (marked by the vertical dashed lines in Fig.~\ref{Fig_Sup3}).  As shown in the inset of this figure, $\Delta\mathcal{S}^{\rm VN}_{\rm ss}$, making the separation between chaotic and regular regimes increasingly pronounced with $S$.

%%%%%%%%%%%%%%%%%%%%% S dependence %%%%%%%%%%%%%%%%%%
%%%%%%%%%%%%%%%%%%%%%%%%%%%%%%%%%%%%%%%%%%%%%%%%%%%%%

\section{Toy model using random matrices}
\label{Sec:Random_matrix_toy_model}

In the main text, we explained that the spectral statistics of the Liouvillian spectrum cannot distinguish transient from steady-state chaos. Here, we present the results for the Liouvillian spectrum of the random matrix toy model introduced in the main text. 
The total Hamiltonian is defined as 
$$\hat{\mathcal{H}} = \hat{\mathcal{H}}_{\rm TD} + \frac{\mu}{\sqrt{N}} \hat{\mathcal{H}}_{\rm I}, $$
where $\hat{\mathcal{H}}_{\rm TD}$ is an $N\times N$ tridiagonal random matrix representing a regular system with Poisson spectral statistics, and $\hat{\mathcal{H}}_{\rm I}$ introduces level repulsion, controlled by the parameter $\mu$. 
Dissipation is incorporated with a jump operator $\sqrt{\gamma}\hat{L}$ where $\hat{L}_{m+1,m}$ are nonzero real random elements for $m = 1, 2,3... $ and $\gamma$ sets the strength of dissipation. The corresponding Liouvillian superoperator is given in Eq.~(2) of the main text.
%%%%%%%%%%%%%%%%%%%%%%%%%%%%%%%%%%%%%%%%%%%%%%%%%%%%
The perturbation Hamiltonian $\hat{\mathcal{H}}_{\rm I}$ is constructed as 
\begin{eqnarray}
	\hat{\mathcal{H}}_{\rm I}(\chi) &=& \hat{\mathcal{P}}(\chi) \hat{\mathcal{H}}_{\rm GOE} \hat{\mathcal{P}}(\chi)\\
	\hat{\mathcal{P}}(\chi) &=& \mathbb{I}-\chi(\ket{u_1}\bra{u_1}+\ket{u_2}\bra{u_2}),
	\label{Projection}
\end{eqnarray}
where $\hat{\mathcal{H}}_{\rm GOE}$ is a random matrix from a Gaussian orthogonal ensemble (GOE) and $\hat{\mathcal{P}}(\chi)$ is the projection operator that suppresses dominant eigenstates $\ket{u_1}$ and $\ket{u_2}$  of the steady-state density matrix $\hat{\rho}_{\rm R}$ of the regular Liouvillian with $\mu=0$ and same jump operator. The parameter $\chi \in [0,1]$ controls the strength of the projection. When $\chi=0$, there is no projection and $\hat{\mathcal{H}}_{\rm I}=\hat{\mathcal{H}}_{\rm GOE}$.

%%%%%%%%%%%%%%%%%%%%% Toy Random matrix %%%%%%%%%%%%%
\begin{figure}[h]
	\includegraphics[width=0.8\columnwidth]{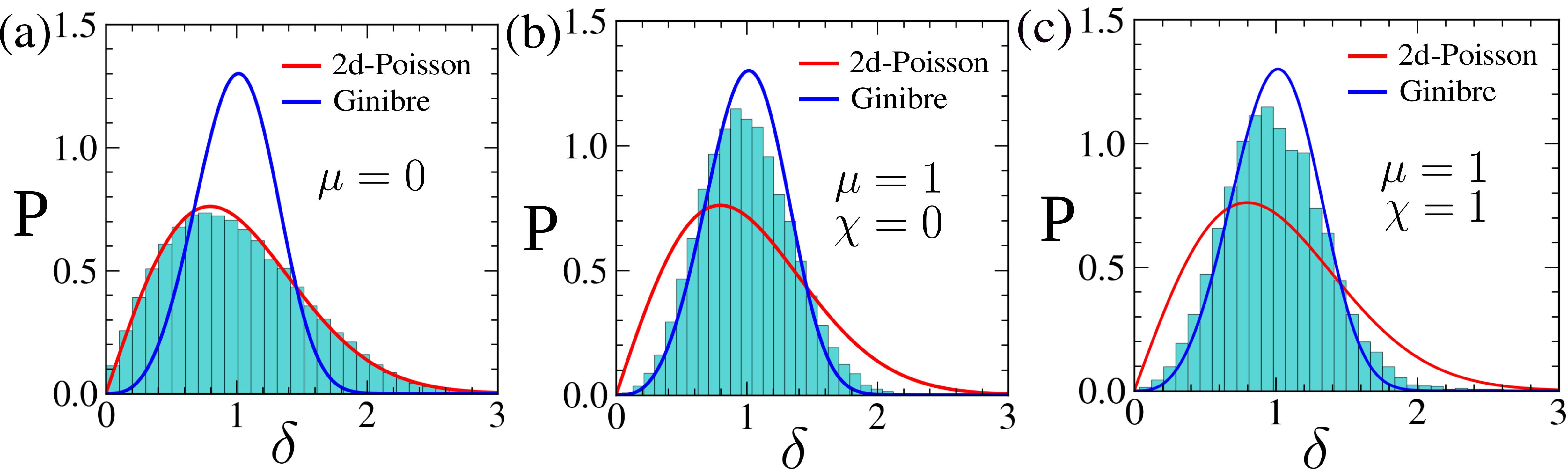}
	\caption{Spectral properties of the Liouvillian for the random matrix toy model. The distribution of nearest neighbor level spacing $\delta$ for the (a) regular Liouvillian with $\mu = 0$ exhibiting 2d-Poisson statistics, (b) Liouvillian with $\mu = 1$ and $\chi = 0$ showing Ginibre statistics, and (c) Liouvillian with $\mu = 1$ and $\chi = 1$ showing again Ginibre statistics. Parameters:  $N = 49, \gamma = 1.0$. }
	\label{Fig_Sup4}
\end{figure}
%%%%%%%%%%%%%%%%%%%%%%%%%%%%%%%%%%%%%%%%%%%%%%%%%%%%

The spectrum of the resulting non-Hermitian Liouvillian superoperator $\mathcal{L}$ is complex and symmetric about the negative real axis. As shown in Figs.~\ref{Fig_Sup4}(a)-(b), when $\chi=0$ and $\mu$ increases from zero, the spectral statistics of $\mathcal{L}$ undergo a crossover from 2d-Poisson   [Fig.~\ref{Fig_Sup4}(a)] to the Ginibre distribution [Fig.~\ref{Fig_Sup4}(b)]. 

At $\mu=1$, the spectrum displays Ginibre statistics regardless of the value of $0<\chi<1$, as seen for $\chi=0$ in Fig.~\ref{Fig_Sup4}(b) and $\chi=1$ in Fig.~\ref{Fig_Sup4}(c). This contrasts with the steady state associated with the zero eigenvalue of the Liouvillian, which shows strong dependence on $\chi$, as shown in Fig.~4(b) of the main text. For large $\chi$, the system retains substantial purity at long times, and the VNE saturates below its maximal value, indicating suppression of steady-state chaos despite early-time signatures of chaos. These results confirm that Ginibre spectral correlations in the Liouvillian spectrum are directly connected to rapid short-time entropy growth, independent of the steady-state behavior, highlighting the limitations of spectral statistics in fully diagnosing chaos in open quantum systems.

%\begin{thebibliography}{99}
%\bibitem{Strogatz}
%S. H. Strogatz, {\it Nonlinear Dynamics and Chaos: With Applications to Physics, Biology, Chemistry, and Engineering}, 2nd ed. (Westview Press, Boulder, CO, 2007).
%	
%\bibitem{Lichtenberg}
%A. J. Lichtenberg and M. A. Lieberman, {\it Regular and Chaotic Dynamics}, 2nd ed. (Springer-Verlag, New York, 1992).
%	
%\bibitem{Hirsch_2016}
%J. Ch\'{a}vez-Carlos, M. A. Bastarrachea-Magnani, S. Lerma-Hern\'{a}ndez, and J. G. Hirsch, Quantum and classical Lyapunov exponents in atom-field interaction systems, Phys. Rev. E {\bf 94}, 022209 (2016).
%	
%\bibitem{Scott_2020}
%K. C. Stitely, A. Giraldo, B. Krauskopf, and S. Parkins, Nonlinear semiclassical dynamics of the unbalanced, open Dicke model, Phys. Rev. Res. {\bf 2}, 033131 (2020).
%	
%\end{thebibliography}

%%%%%%% THE END %%%%%%%
\end{document}